\documentclass[twocolumn,showpacs]{revtex4}
\usepackage{graphicx,subfigure}
\begin{document}
\title{Alternative criterion for two-dimensional wrapping percolation}
\author{Hongting Yang}
\affiliation{School of Science, Wuhan University of Technology, Wuhan 430070, P.R. China}
\date{\today}

\begin{abstract}
Based on the differences between a spanning cluster and a wrapping cluster, an alternative criterion for testing wrapping
percolation is provided for two-dimensional lattices. By following the Newman-Ziff method, the finite size scaling of
estimates for percolation thresholds are given. The results are consistent with those from Machta's method.
\end{abstract}

\maketitle

\section{introduction}
In a square lattice each site is independently either occupied with probability $p$, or empty with probability $1-p$. A
cluster is a group of occupied (nearest) neighbor sites~\cite{stauffer}. With more and more sites being occupied, clusters
grow larger and larger. For the square lattice with free boundaries, once a cluster grows large enough to touch the two
opposite boundaries, the spanning percolation has occurred. This cluster is called as a spanning cluster. While for the
lattice with periodic boundary conditions, it is somewhat more difficult to detect a wrapping cluster, which wraps around
the lattice.

Since the very early work on three dimensional polymers~\cite{flory1941}, percolation has found a variety of uses in many
fields. Typically in the study of networks~\cite{son2012,bizhani2011,agliari2011,sergey2010}, people keep casting great interests.
Recent studies of percolation in physics mainly involve various functional materials or components, such as
optical lattices~\cite{KV2007}, magnetic materials or ordinary materials~\cite{HV2006,ortuno2011,lois2009}, and
nanocomposites~\cite{ofir2007,LM2008,stevens2008}. Very recently, the study on explosive percolation arose
``explosive'' interests~\cite{araujo2011,costa2010,araujo2010,radicchi2010}. Many other interesting works on percolation theory
and its wide applications are collected in some books~\cite{stauffer,sahimi1994,hunt2009}.

In all these studies of percolation, a naive question is that how to tell the onset of percolation. The percolation
threshold (or critical probability) $p_c$ is the value of $p$ for which a spanning cluster (in the case of free boundaries)
or a wrapping cluster (periodic boundary conditions) appears for the first time~\cite{NZ2001,PM2003}. For a periodic
lattice, once a wrapping cluster appears, say along the x-direction, if we cut the lattice along y-direction to let the
x-boundaries be open, there exists (at least) one spanning cluster. In turn, when a spanning cluster first appears on a
free boundaries lattice, if we connect the two opposite open boundaries to let it be a periodic lattice, the wrapping
cluster does not necessarily appear. Therefore, wrapping percolation always happens later than spanning percolation for the
lattices with the same linear dimension $L$ but with different boundary conditions. The difference between a wrapping
cluster and a spanning cluster gives us a clue to build an alternative criterion for wrapping percolation. Once a spanning
cluster occurs, we amalgamate those clusters (excluding the spanning cluster) touching the boundaries, build connected
relations between the spanning cluster and the amalgamated clusters. If finally, two ends of the spanning cluster connect
to each other via an amalgamated cluster, then a wrapping cluster appears. The data processing after obtaining a wrapping
cluster follows the Newman-Ziff algorithm~\cite{NZ2001} exactly.

The prevalent criterion for wrapping percolation introduced by Machta {\em et al.}~\cite{machta96} has been described in
detail for bond percolation in Newman-Ziff algorithm. It can be easily extend to site percolation, once the displacements
of neighboring sites to the same root site differ by an amount other than zero or one lattice spacing, the cluster wrapping
has occurred. Taking site percolation as an example, the core idea of Newman-Ziff algorithm is that, starting with an empty
lattice, a percolation state can be realized simply by adding sites one by one to the lattice, a sample state with $n+1$
occupied sites is achieved by adding one extra randomly chosen site to a sample state with $n$ sites. An important
technique in the Newman-Ziff algorithm is the application of binomial distribution. Taking the number of occupied sites $n$
as ``energy'', if we can get a set of measurements $\{Q_n\}$ in the microcanonical ensemble, then the observable $Q(p)$ in
the canonical ensemble is given by
\begin{equation}
 Q(p)=\sum_{n=0}^N{N\choose n}p^n(1-p)^{N-n}Q_n.
\end{equation}
Such observable $Q$ can be probability of cluster wrapping, mean cluster size, and correlation length etc. The main
advantage of this technique is that, the continuous observable $Q(p)$ for all $p$ can be determined from discrete $N+1$
values of $Q_n$.

For convenience, we use the same symbols as in Newman-Ziff algorithm. There are four types of probability $R_L(p)$ of
cluster wrapping on periodic square lattice of $L\times L$ sites. $R_L^{(e)}$ is the probability of cluster wrapping along
either the horizontal or vertical directions, or both; $R_L^{(1)}$, around one specified axis but not the other axis;
$R_L^{(b)}$, in both horizontal and vertical directions; $R_L^{(h)}$ and $R_L^{(v)}$, around the horizontal and vertical
directions, respectively. For square lattice, $R_L^{(h)}=R_L^{(v)}$. These four wrapping probabilities satisfy the
equations
\begin{eqnarray}
  R_L^{(b)} &=& R_L^{(e)}-2R_L^{(1)}, \\
  R_L^{(h)} &=& R_L^{(e)}-R_L^{(1)},
\end{eqnarray}
from which we get the values of $R_L^{(b)}$ and $R_L^{(h)}$ by measuring only the values of $R_L^{(e)}$ and $R_L^{(1)}$, or
vice versa. Given the exact value of $R_\infty(p_c)$, the solution $p$ of the equation
\begin{equation}
  R_L(p)=R_\infty(p_c)
\end{equation}
gives a very good estimator for $p_c$. The solution $p$ converges to $p_c$ according to
\begin{equation}
  p-p_c\sim L^{-11/4}.
\end{equation}
In this way we get the corresponding values of $p_c$ from $R_\infty^{(e)}(p_c)$, $R_\infty^{(b)}(p_c)$ and
$R_\infty^{(h)}(p_c)$ respectively. Since the wrapping probability $R_L^{(1)}(p)$ is nonmonotonic, we use the position of
its maximum to estimate $p_c$, instead of the value of $R_\infty^{(1)}(p_c)$.

The CPU time $T_L$ is related to the statistical errors $\sigma_{p_c}$ according to $\sigma_{p_c}\sim T_L^{-1/2}L^{1/4}$.
For finite $L$ and a specific algorithm, $T_L$ depends only on $n$, which represents the number of runs of the algorithm,
instead of the number of occupied sites. To keep the same statistical errors on systems of different size, we take
appropriate $n$ to fulfill $T_L\sim \sqrt{L}$.

In Sec. \ref{algorithm} we describe our algorithm. In Sec. \ref{results} we give the values of $p_c$ from the algorithm,
and compare it with those from Machta's method. In Sec. \ref{conclusions} we give our conclusions.

\section{the algorithm} \label{algorithm}
\begin{figure}[h]
\centering \subfigure[]{\label{alg1}\includegraphics[width=0.15\textwidth]{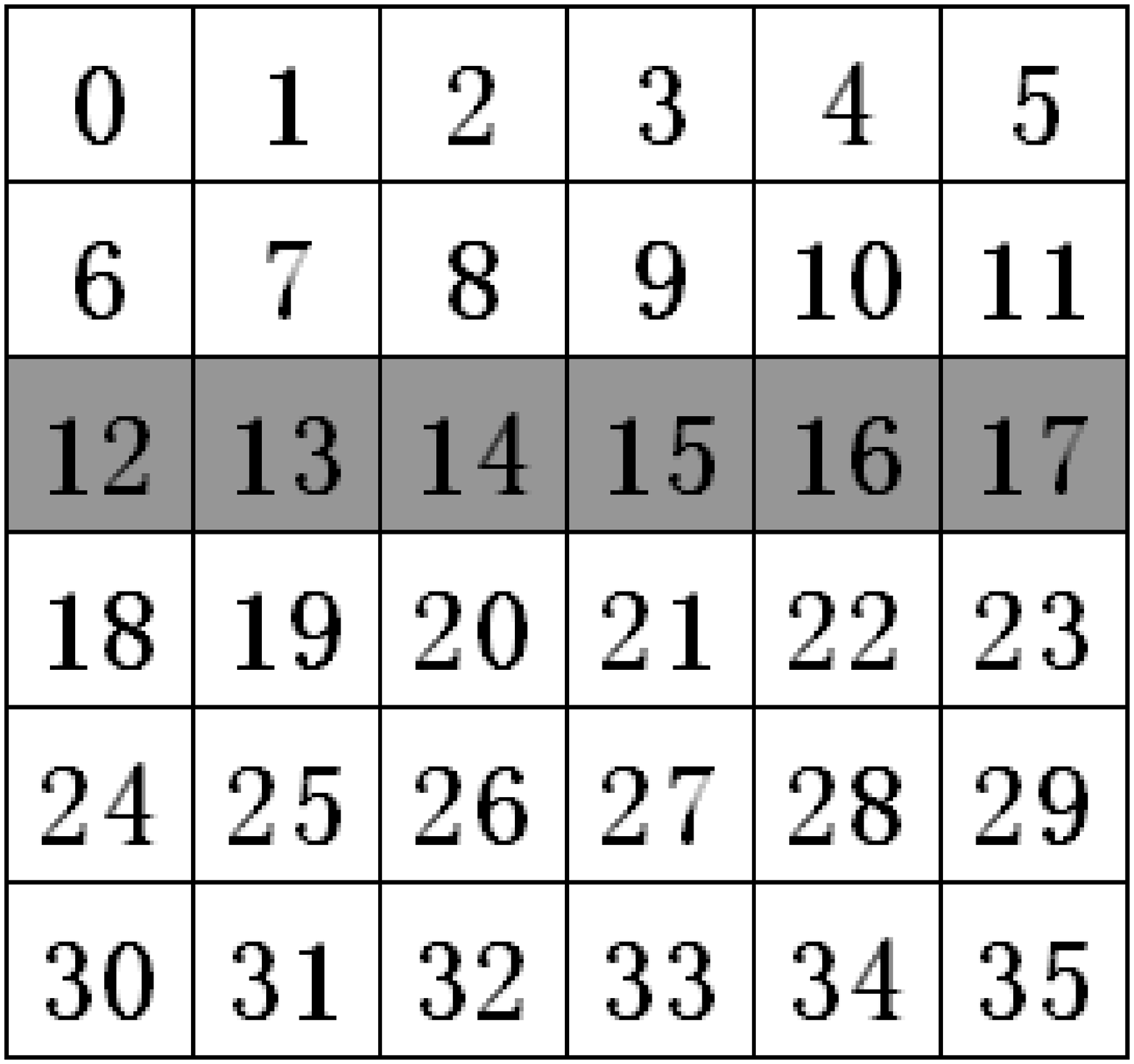}} \hspace{-5pt}
\subfigure[]{\label{alg2}\includegraphics[width=0.15\textwidth]{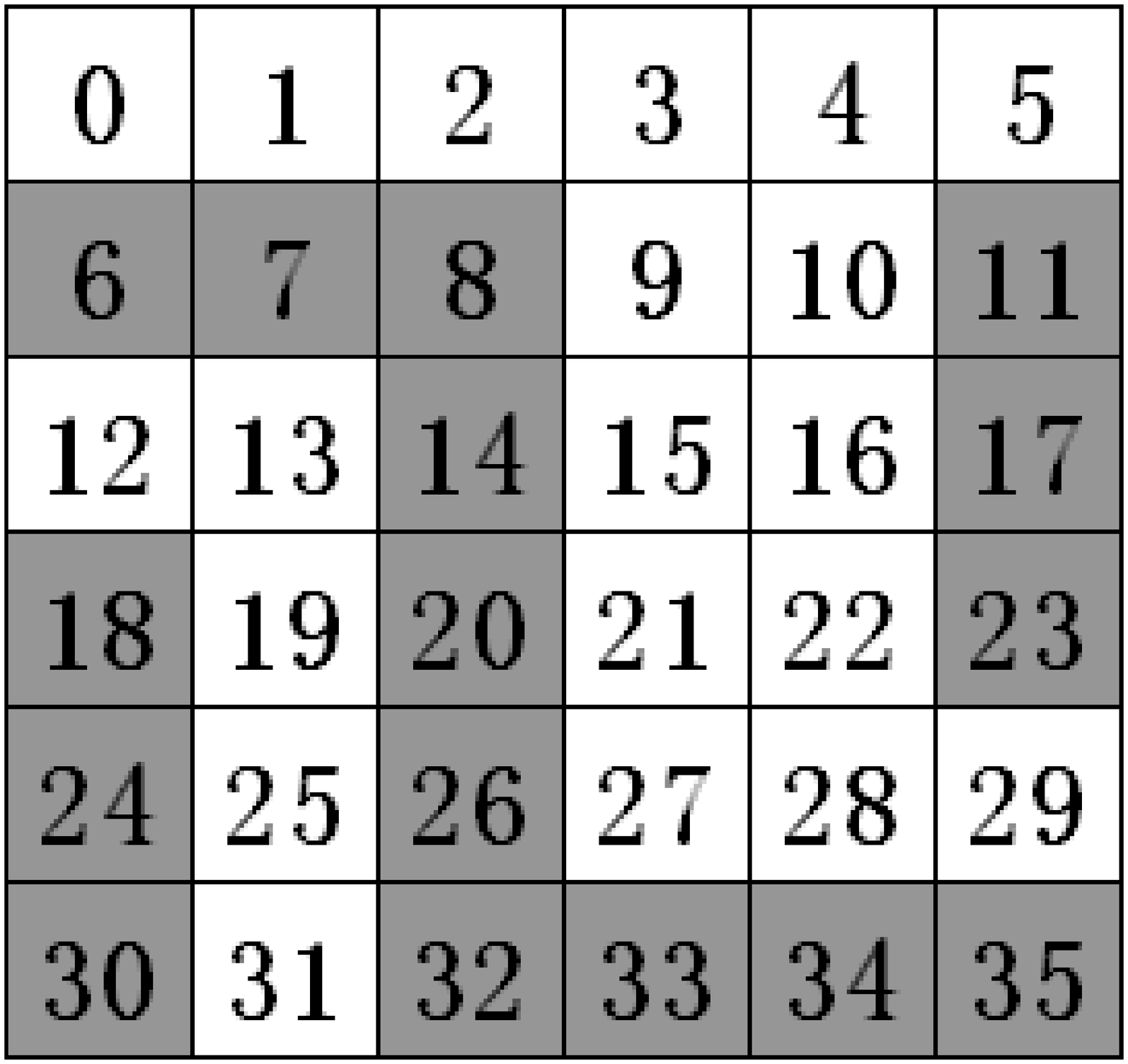}} \hspace{-5pt}
\subfigure[]{\label{alg3}\includegraphics[width=0.15\textwidth]{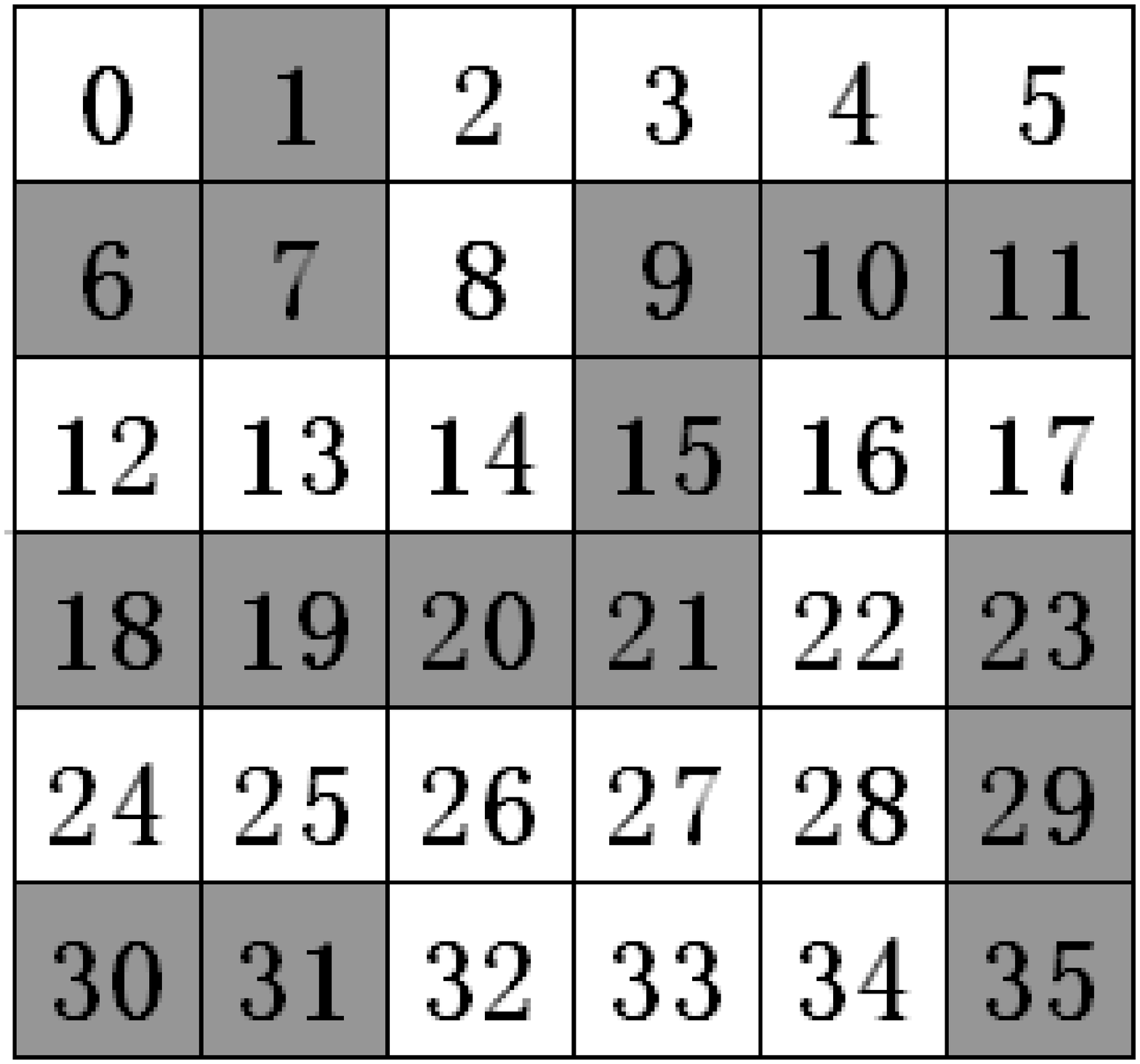}} \caption{Examples of wrapping clusters with
occupied sites shaded. Pairs of clusters (excluding the spanning cluster) are merged at the first step of scanning the
occupied quasi-neighbour sites, {\em e.g.} (18, 23) in (b) and (30, 35) in (c).  In (a), the occupied quasi-neighbour sites
(12, 17) belong to the same cluster. In (b) or (c), two ends of the spanning cluster are connected via a merged cluster.}
\end{figure}

Since our criterion for wrapping percolation originates from the difference in wrapping cluster and spanning cluster, the
central task of the method is how to handle those clusters touching the boundaries. Each run of the algorithm starts from a
lattice with periodic boundary conditions. In the following, we take the site wrapping percolation along x-direction as an
example. The whole algorithm can be separated into two parts, the spanning process and the wrapping process. In the
spanning process, we check the states of a newly occupied site and its (nearest) neighbour sites previously occupied. If
one site and its neighbour happen to be on the left boundary and right boundary respectively, we call them as a pair of
occupied ``quasi-neighbour'' sites. Along x-direction, if we shift the periodic boundary conditions to the open boundary
conditions, two sites of any pair of quasi-neighbour sites are no longer neighboring to each other. In Fig.~\ref{alg1}, six
pairs of quasi-neighbour sites on the square lattice $6\times6$ are (0,5), (6,11), (12,17), (18,23), (24,29), (30,35). Two
occupied quasi-neighbor sites belong to their respective clusters. Therefore, pairs of occupied quasi-neighbour sites
correspond to pairs of their respective clusters. No implementations are required to these pairs of clusters or these
occupied quasi-neighbour sites. For non-quasi-neighbour sites, if the two sites point to the same root site (belong to the
same cluster), we need do nothing; otherwise, if the two sites belong to different clusters, we must merge them into a
single cluster. We do exactly what we usually do~\cite{gould2006} before the appearance of a spanning cluster~\footnote{For
the lattice with open boundary conditions along x-direction, but with periodic boundary conditions along y-direction, it is
possible there are two or more spanning clusters along x-direction. In the text, the spanning cluster always indicates that
one right visited, all other touching-boundary clusters are grouped into non-spanning clusters whether they are spanning or
not.} on a square lattice. In other words, the periodic boundary conditions along x-direction are suppressed temporarily in
the spanning process.

Once a spanning cluster occurs, we turn to the wrapping process immediately, which aims to find a wrapping cluster.
Obviously, this process is the core of the present algorithm. A spanning cluster has at least two ends (left end and right
end), which touch the left boundary and the right boundary respectively. When the periodic boundary conditions along
x-direction are recovered, by across the boundaries, the two ends of the spanning cluster could be connected, which implies
the appearance of a wrapping cluster. This process can be finished in three steps.

In the first step, we check the pairs of clusters touching the boundaries by scanning the occupied quasi-neighbour sites
row-by-row. If two clusters that a pair of occupied quasi-neighbour sites respectively belong to own different root sites
and neither of them is the spanning cluster, we merge them into a single cluster. The root site of any of them can be
chosen as the root site of the merged cluster. After the implementations on all the non-spanning clusters touching the
boundaries, what we do next is to build relations between the spanning cluster and the (merged) non-spanning clusters
touching the boundaries. We may meet three interesting cases when we scan pairs of clusters touching the boundaries once
more. The simplest case is that the two clusters are the same spanning cluster (see Fig.~\ref{alg1} as an example). We
simply add a pointer from the spanning cluster to itself. The second case is that, a non-spanning cluster does not possess
a pointer pointing to the spanning cluster, we could add a pointer from the former to the latter, and label the boundary
touched by the non-spanning cluster. For the third case, a non-spanning cluster already has a pointer pointing to the
spanning cluster and previously touches the opposite boundary, we add a pointer from the spanning cluster to the
non-spanning cluster. Two examples for the third case are showed in Figs.~\ref{alg2} and \ref{alg3} respectively. For other
cases, we need do nothing further. In the final step, a simple while statement is used to check whether the spanning
cluster points to itself or not by following the pointers added in the second step.

Thus the wrapping process of our algorithm can be summarized as follows.
\begin{description}
  \item[(1)] Amalgamate pairs of non-spanning touching-boundary clusters.
  \item[(2)] Add pointers from the amalgamated non-spanning touching-boundary clusters to the spanning cluster, or in turn,
            add a pointer from the spanning cluster to a non-spanning touching-boundary cluster.
  \item[(3)] By following a succession of pointers added above, check if we can get from the spanning-cluster to itself.
\end{description}

If no wrapping cluster appears, we occupy one additional site on the previous lattice and repeat the spanning process and
the wrapping process. Since we aim to check the reliability of our algorithm via the calculation of percolation threshold
only, we halt the algorithm once the percolation along both x and y directions are detected. The program for our algorithm
written in C is available online~\cite{y1SM}.

\section{the results} \label{results}
\begin{figure}[h]
\centering \subfigure[]{\label{pc}\includegraphics[width=0.5\textwidth]{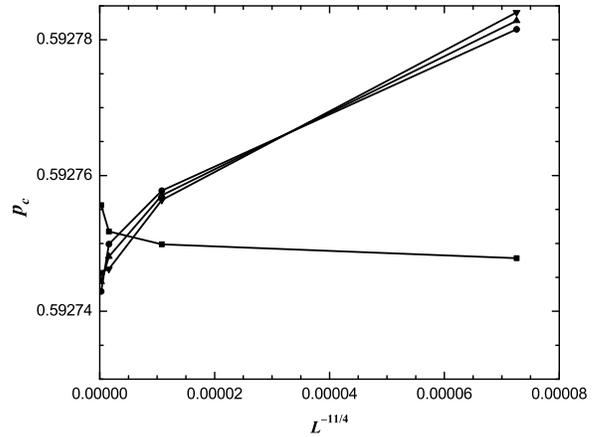}}
\subfigure[]{\label{pcM}\includegraphics[width=0.5\textwidth]{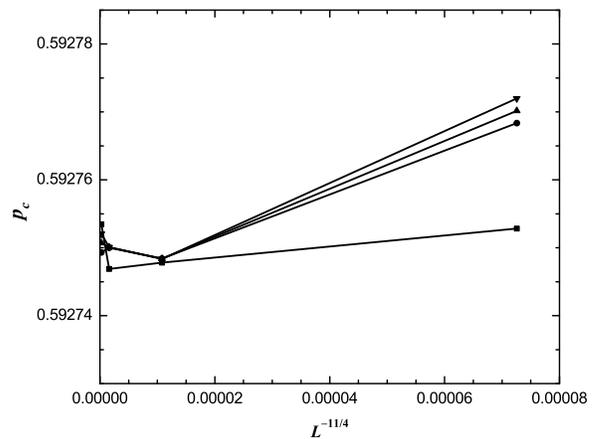}}
\caption{\label{2pc}The values of $p_c$ on square
lattices of $L\times L$ obtained from the probabilities of cluster wrapping along one axis but not the other (solid
squares), both axes (solid circles), one axis (solid upward-pointing triangles), and either axis (solid downward-pointing
triangles). (a) Our method; (b) Machta's method.}
\end{figure}

All computations are implemented on a desktop PC with CPU clock speed 2.6 GHz and memory 1.96 GB. We fix $n$, runs of the
algorithm, to $2\times10^6$ for the lattice with $L=256$. The computation takes CPU time about 37 hours. The other values
of $n$ for $L$ equal to 128, 64 and 32 are respectively chosen to ensure $T_L\sim \sqrt{L}$. The total $n$ is about
$2.7\times10^8$. The finite size scaling of $p_c$ for square lattice $L\times L$ is showed in Fig.~\ref{pc}, and the values
of $p_c$ obtained from four different probabilities $R_L$ are listed in Table~\ref{pcTable}.
\begin{table}[h]
  \centering
  \caption{The values of $p_c$ for infinite lattice. Here, {\sf P} and {\sf M} represents the present method and
   Machta's method respectively; $p_c^{(1)}$, $p_c^{(b)}$, $p_c^{(h)}$ and $p_c^{(e)}$ corresponds respectively to the
   wrapping probabilities $R_L^{(1)}$, $R_L^{(b)}$, $R_L^{(h)}$ and $R_L^{(e)}$.}\label{pcTable}
\begin{tabular}{c|c|c|c|c}
  \hline
       & $p_c^{(1)}$ & $p_c^{(b)}$ & $p_c^{(h)}$ & $p_c^{(e)}$ \\ \hline
  {\sf P}  & 0.5927528(16) & 0.5927480(31) & 0.5927475(23) & 0.5927471(19) \\ \hline
  {\sf M}  & 0.5927493(23) & 0.5927482(14) & 0.5927487(18) & 0.5927491(22) \\
  \hline
\end{tabular}
\end{table}

To check the reliability of our method, and as a comparison, a similar computation within the framework of Machta's method
is implemented. For $L=256$, $n$ is also taken as $2\times10^6$, CPU time is about 4.5 hours, which is only one eighth of
that in our method. In other words, the computation time in units of hours is $T_M= 0.28\sqrt{L}$ for Machta's method,
while for our method $T_P= 2.3\sqrt{L}$. The other values of $n$ corresponding to $L$ are chosen by reference to this short
CPU time, and the total $n$ in Machta's method is about $8\times 10^7$. In comparison with the simulation of more than
$7\times 10^9$ separate samples in the work of Newman-Ziff, the statistical errors in this work are larger. The finite size
scaling and the results of $p_c$ are respectively showed in Fig.~\ref{pcM} and Table~\ref{pcTable}.

Besides the difference in computation time, there are tiny differences in the results. Obviously, from Table~\ref{pcTable},
one can see that the data of $p_c^{(1)}$ and $p_c^{(e)}$ are better (smaller errors) than the data of $p_c^{(b)}$ and
$p_c^{(h)}$ in our method, while in the Machta's method, on the contrary, the data of $p_c^{(b)}$ and $p_c^{(h)}$ are
better. This difference possibly arises from the larger values of $p_c^{(e)}$ (and therefore larger values of $p_c^{(b)}$
and $p_c^{(h)}$) at small $L$ in our method, which can be seen in Fig.~\ref{2pc}. Except these differences, our results
coincide with that from Machta's method quite well.

\section{the conclusions} \label{conclusions}
A wrapping cluster appears, if two ends (touching two opposite boundaries respectively) of a spanning cluster are connected
to each other by across the boundaries. The results of percolation threshold in our method are as good as that from
Machta's method, and are consistent with the published estimates of the square site percolation
threshold~\cite{deng2005,feng2008,lee2008}. In comparison with Machta's method, although our method is not competitive in
computation time, it provides a unified method for wrapping percolation and spanning percolation. This work helps us to
choose an appropriate method for the further study of some kind of spatially correlated percolation model~\cite{yang2009},
where the percolation thresholds have not yet been well determined.

The lengthy computation time in our method mainly comes from two aspects of our specific algorithm. In the spanning
process, the clusters are amalgamated by suppressing the periodic boundary conditions along x and y directions
respectively; after a spanning cluster appears (either along x or y direction), a wrapping process for testing a wrapping
cluster is implemented. While in the Machta's method, a wrapping cluster could be tested in the process of merging clusters
on a square lattice with full periodic boundary conditions.

Either although the values of percolation threshold could be obtained from spanning percolation on a lattice with open
boundary conditions or from wrapping percolation on a lattice with periodic boundary conditions, different boundary effects
are definitely covered in the results. With our method, it is possible to give a direct computation of different boundary
effects.

The present algorithm differs from the Newman-Ziff algorithm only in the criterion for wrapping percolation. Without
question, the former can be extended to the calculation of $p_c$ for three-dimensional percolation on the cubic lattice in
the same way as that of the latter.

\section{acknowledgments}
The author would like to thank Prof. Stephan Haas for active comments. The recommendations and criticisms of the referees
are highly appreciated to improve the present report.


\begin{thebibliography}{99}
\bibitem{stauffer} D. Stauffer and A. Aharony, {\em Introduction to Percolation Theory}, 2nd ed. (Taylor \& Francis, London, 1992).
\bibitem{flory1941} P. J. Flory, J. Am. Chem. Soc. {\bf 63}, 3083 (1941).
\bibitem{son2012} S. -W. Son, G. Bizhani, C. Christensen, P. Grassberger and M. Paczuski, EPL {\bf 97}, 16006(2012).
\bibitem{bizhani2011} G. Bizhani, P. Grassberger and M. Paczuski, Phys. Rev. E {\bf 84}, 066111 (2011).
\bibitem{agliari2011} E. Agliari, C. Cioli and E. Guadagnini, Phys. Rev. E {\bf 84}, 031120 (2011).
\bibitem{sergey2010} S. V. Buldyrev, R. Parshani, G. Paul, H. E. Stanley and S. Havlin,  Nature {\bf 464}, 1025 (2010).
\bibitem{KV2007} Y.V. Kartashov, V.A. Vysloukh and L. Torner, Opt. Exp. {\bf 15}, 12409 (2007).
\bibitem{HV2006} J.A. Hoyos and T. Vojta, Phys. Rev. B {\bf 74}, 140401 (2006).
\bibitem{ortuno2011} M. Ortu\~{n}o, A. M. Somoza, V. V. Mkhitaryan and M. E. Raikh, Phys. Rev. B {\bf 84}, 165314 (2011).
\bibitem{lois2009} G. Lois, J. Blawzdziewicz and C. S. O'Hern, Phys. Rev. Lett. {\bf 102}, 015702 (2009).
\bibitem{ofir2007} A. Ofir, S. Dor, L. Grinis, A. Zaban, T. Dittrich and J. Bisquert, J. Chem. Phys. {\bf 128}, 064703 (2008).
\bibitem{LM2008} C. Lu and Y.-W. Mai, J. Mater. Sci. {\bf 43}, 6012 (2008).
\bibitem{stevens2008} D.R. Stevens, L.N. Downen and L.I. Clarke, Phys. Rev. B {\bf 78}, 235425 (2008).
\bibitem{araujo2011} N. A. M. Ara\'ujo, J. S. Andrade Jr, R. M. Ziff and H. J. Herrmann, Phys. Rev. Lett. {\bf 106}, 095703 (2011).
\bibitem{costa2010} R. A. da Costa, S. N. Dorogovtsev, A. V. Goltsev and J. F. F. Mendes, Phys. Rev. Lett. {\bf 105}, 255701 (2010).
\bibitem{araujo2010} N. A. M. Ara\'ujo and H. J. Herrmann, Phys. Rev. Lett. {\bf 105}, 035701 (2010).
\bibitem{radicchi2010} F. Radicchi and S. Fortunato, Phys. Rev. E {\bf 81}, 036110 (2010).
\bibitem{sahimi1994} M. Sahimi, {\em Applications of Percolation Theory}, Taylor \& Francis, Bristol, MA, 1994.
\bibitem{hunt2009} A. G. Hunt and R. Ewing, {\em Percolation Theory for Flow in Porous Media}, Lect. Notes Phys. 771, 2nd edition,
                   Springer, Berlin Heidelberg, 2009.
\bibitem{NZ2001} M. E. J. Newman and R. M. Ziff, Phys. Rev. E {\bf 64}, 016706 (2001).
\bibitem{PM2003} G. Pruessner and N. R. Moloney, J. Phys. A {\bf 36}, 11213 (2003).
\bibitem{machta96} J. Machta, Y. S. Choi, A. Lucke, T. Schweizer, and L. M. Chayes, Phys. Rev. E {\bf 54}, 1332 (1996).
\bibitem{gould2006} H. Gould, J. Tobochnik and W. Christian, {\em An Introduction to Computer Simulation Methods}, 3rd ed.
                  (Addison-Wesley, Reading, MA, 2006), p.~468.
\bibitem{y1SM} See Supplemental Material at [URL inserted here by publisher] for the program of our wrapping percolation algorithm.
\bibitem{deng2005} Y. Deng and H. W. J. Bl\"{o}te, Phys. Rev. E {\bf 72}, 016126 (2005)
\bibitem{feng2008} X. Feng, Y. Deng and H. W. J. Bl\"{o}te, Phys. Rev. E {\bf 78}, 031136 (2008)
\bibitem{lee2008} M. J. Lee, Phys. Rev. E {\bf 78}, 031131 (2008).
\bibitem{yang2009} H. Yang, W. Zhang, N. Bray-Ali, and S. Haas, arXiv:0908.0104v2.
\end{thebibliography}
\end{document}